\documentstyle[twoside,psfig]{article}
\catcode`\@=11
\long\def\@makefntext#1{
\protect\noindent \hbox to 3.2pt {\hskip-.9pt
$^{{\eightrm\@thefnmark}}$\hfil}#1\hfill}               

\def\@makefnmark{\hbox to 0pt{$^{\@thefnmark}$\hss}}    

\def\ps@myheadings{\let\@mkboth\@gobbletwo
\def\@oddhead{\hbox{}
\rightmark\hfil\eightrm\thepage}
\def\@oddfoot{}\def\@evenhead{\eightrm\thepage\hfil
\leftmark\hbox{}}\def\@evenfoot{}
\def\sectionmark##1{}\def\subsectionmark##1{}}

\oddsidemargin=\evensidemargin
\newcounter{sectionc}\newcounter{subsectionc}\newcounter{subsubsectionc}
\renewcommand{\section}[1] {\vspace{12pt}\addtocounter{sectionc}{1}
\setcounter{subsectionc}{0}\setcounter{subsubsectionc}{0}\noindent
        {\tenbf\thesectionc. #1}\par\vspace{5pt}}
\renewcommand{\subsection}[1] {\vspace{12pt}\addtocounter{subsectionc}{1}
      \setcounter{subsubsectionc}{0}\noindent
      {\bf\thesectionc.\thesubsectionc.{\kern1pt \bfit #1}}\par\vspace{5pt}}
\renewcommand{\subsubsection}[1]
      {\vspace{12pt}\addtocounter{subsubsectionc}{1}
      \noindent{\tenrm\thesectionc.\thesubsectionc.\thesubsubsectionc.
      {\kern1pt \tenit #1}}\par\vspace{5pt}}
\newcommand{\nonumsection}[1] {\vspace{12pt}\noindent{\tenbf #1}
        \par\vspace{5pt}}

\newcounter{appendixc}
\newcounter{subappendixc}[appendixc]
\newcounter{subsubappendixc}[subappendixc]
\renewcommand{\thesubappendixc}{\Alph{appendixc}.\arabic{subappendixc}}
\renewcommand{\thesubsubappendixc}
        {\Alph{appendixc}.\arabic{subappendixc}.\arabic{subsubappendixc}}

\renewcommand{\appendix}[1] {\vspace{12pt}
        \refstepcounter{appendixc}
        \setcounter{figure}{0}
        \setcounter{table}{0}
        \setcounter{lemma}{0}
        \setcounter{theorem}{0}
        \setcounter{corollary}{0}
        \setcounter{definition}{0}
        \setcounter{equation}{0}
        \renewcommand{\thefigure}{\Alph{appendixc}.\arabic{figure}}
        \renewcommand{\thetable}{\Alph{appendixc}.\arabic{table}}
        \renewcommand{\theappendixc}{\Alph{appendixc}}
        \renewcommand{\thelemma}{\Alph{appendixc}.\arabic{lemma}}
        \renewcommand{\thetheorem}{\Alph{appendixc}.\arabic{theorem}}
        \renewcommand{\thedefinition}{\Alph{appendixc}.\arabic{definition}}
        \renewcommand{\thecorollary}{\Alph{appendixc}.\arabic{corollary}}
        \renewcommand{\theequation}{\Alph{appendixc}.\arabic{equation}}
        \noindent{\tenbf Appendix \theappendixc #1}\par\vspace{5pt}}
\newcommand{\subappendix}[1] {\vspace{12pt}
        \refstepcounter{subappendixc}
        \noindent{\bf Appendix \thesubappendixc. {\kern1pt \bfit #1}}
        \par\vspace{5pt}}
\newcommand{\subsubappendix}[1] {\vspace{12pt}
        \refstepcounter{subsubappendixc}
        \noindent{\rm Appendix \thesubsubappendixc. {\kern1pt \tenit #1}}
        \par\vspace{5pt}}

\newcommand{\smalllineskip}{\baselineskip=10pt}

\def\eightcirc{
\begin{picture}(0,0)
\put(4.4,1.8){\circle{6.5}}
\end{picture}}
\def\eightcopyright{\eightcirc\kern2.7pt\hbox{\eightrm c}}

\def\abstracts#1#2#3{{
        \centering{\begin{minipage}{4.5in}\baselineskip=10pt\footnotesize
        \parindent=0pt #1\par
        \parindent=15pt #2\par
        \parindent=15pt #3
        \end{minipage}}\par}}

\renewenvironment{thebibliography}[1]
        {\frenchspacing
         \ninerm\baselineskip=11pt
         \begin{list}{\arabic{enumi}.}
        {\usecounter{enumi}\setlength{\parsep}{0pt}
         \setlength{\leftmargin 12.7pt}{\rightmargin 0pt} 
         \setlength{\itemsep}{0pt} \settowidth
        {\labelwidth}{#1.}\sloppy}}{\end{list}}

\newcounter{itemlistc}
\newcounter{romanlistc}
\newcounter{alphlistc}
\newcounter{arabiclistc}

\newcommand{\fcaption}[1]{
        \refstepcounter{figure}
        \setbox\@tempboxa = \hbox{\footnotesize Fig.~\thefigure. #1}
        \ifdim \wd\@tempboxa > 5in
           {\begin{center}
        \parbox{5in}{\footnotesize\smalllineskip Fig.~\thefigure. #1}
            \end{center}}
        \else
             {\begin{center}
             {\footnotesize Fig.~\thefigure. #1}
              \end{center}}
        \fi}

\newcommand{\tcaption}[1]{
        \refstepcounter{table}
        \setbox\@tempboxa = \hbox{\footnotesize Table~\thetable. #1}
        \ifdim \wd\@tempboxa > 5in
           {\begin{center}
        \parbox{5in}{\footnotesize\smalllineskip Table~\thetable. #1}
            \end{center}}
        \else
             {\begin{center}
             {\footnotesize Table~\thetable. #1}
              \end{center}}
        \fi}

\def\@citex[#1]#2{\if@filesw\immediate\write\@auxout
        {\string\citation{#2}}\fi
\def\@citea{}\@cite{\@for\@citeb:=#2\do
        {\@citea\def\@citea{,}\@ifundefined
        {b@\@citeb}{{\bf ?}\@warning
        {Citation `\@citeb' on page \thepage \space undefined}}
        {\csname b@\@citeb\endcsname}}}{#1}}

\newif\if@cghi
\def\cite{\@cghitrue\@ifnextchar [{\@tempswatrue
        \@citex}{\@tempswafalse\@citex[]}}
\def\citelow{\@cghifalse\@ifnextchar [{\@tempswatrue
        \@citex}{\@tempswafalse\@citex[]}}
\def\@cite#1#2{{$\null^{#1}$\if@tempswa\typeout
        {IJCGA warning: optional citation argument
        ignored: `#2'} \fi}}

\def\@refcitex[#1]#2{\if@filesw\immediate\write\@auxout
        {\string\citation{#2}}\fi
\def\@citea{}\@refcite{\@for\@citeb:=#2\do
        {\@citea\def\@citea{, }\@ifundefined
        {b@\@citeb}{{\bf ?}\@warning
        {Citation `\@citeb' on page \thepage \space undefined}}
        \hbox{\csname b@\@citeb\endcsname}}}{#1}}

\def\@refcite#1#2{{#1\if@tempswa\typeout
        {IJCGA warning: optional citation argument
        ignored: `#2'} \fi}}

\def\refcite{\@ifnextchar[{\@tempswatrue
        \@refcitex}{\@tempswafalse\@refcitex[]}}

\def\pmb#1{\setbox0=\hbox{#1}
        \kern-.025em\copy0\kern-\wd0
        \kern.05em\copy0\kern-\wd0
        \kern-.025em\raise.0433em\box0}

\def\fnt#1#2{\footnotetext{\kern-.3em
        {$^{\mbox{\scriptsize #1}}$}{#2}}}

\def\fpage#1{\begingroup
\voffset=.3in
\thispagestyle{empty}\begin{table}[b]\centerline{\footnotesize #1}
        \end{table}\endgroup}

\def\runninghead#1#2{\pagestyle{myheadings}
\markboth{{\protect\footnotesize\it{\quad #1}}\hfill}
{\hfill{\protect\footnotesize\it{#2\quad}}}}
\font\tenrm=cmr10
\font\tenit=cmti10
\font\tenbf=cmbx10
\font\bfit=cmbxti10 at 10pt
\font\ninerm=cmr9

\font\eightrm=cmr8

\def\qed{\hbox{${\vcenter{\vbox{                      
   \hrule height 0.4pt\hbox{\vrule width 0.4pt height 6pt
   \kern5pt\vrule width 0.4pt}\hrule height 0.4pt}}}$}}

\begin{document}

\runninghead{Yu. Baurov}
{Structure of physical space$\ldots$}

\thispagestyle{empty}\setcounter{page}{1}
\vspace*{0.88truein}
\fpage{1}

\centerline{\bf STRUCTURE OF PHYSICAL SPACE}
\centerline{\bf AND NATURE OF ELECTROMAGNETIC FIELD}

\vspace*{0.035truein}

\vspace*{0.37truein}
\centerline{\footnotesize Yuri A. Baurov}

\centerline{\footnotesize \it
Central Research Institute of Machine Building}
\baselineskip=10pt
\centerline{\footnotesize \it
Pionerskaya, 4}
\baselineskip=10pt
\centerline{\footnotesize \it
141070, Korolyov, Moscow Region, Russia}

\baselineskip 5mm

\vspace*{0.21truein}

\abstracts{ In the article, on the basis of the author's model of formation
of the observable physical space $R_3$ in the process of dynamics
of special discrete one-dimensional vectorial objects, {\it byuons},
while minimizing their potential energy of interaction in the
one-dimensional space $R_1$ formed by them, and based on the Weyl's
geometrical model of the electromagnetic field, the nature of
this field consisting in a suitable variation of byuon
interaction periods in the space $R_1$ , is shown.}{}{}

\bigskip

$$$$

\section{\bf Introduction }
\vskip10pt
A wealth of works, beginning from antiquity
(Aristoteles\cite{1}, "The Principia" by Euclides, Democritus) and
ending with authors of 20th century, are dedicated to the
structure of space and time, to the physical sense of these
fundamental concepts, and their properties. For brevity sake, we
will not give an extensive bibliography on the subject but
advise the reader to acquaint himself only with two monographs
\cite{2,3}, in the first of which a physical and philosophical
comprehension of hundreds of works on this problem is presented,
and in the second, an ingenious work of a known russian
physicist D.I.Blokhintsev, difficulties encountered in
describing the world of elementary particles when creating a
modern space theory, are shown.  In all existing works on
quantum field theory and physics of elementary particles, the
space in which elementary processes occur, as a rule, is given
one way or another. Further the action $S_1$ is written in the
given space in terms of the Lagrangian function density for the
fields and objects considered, and equations of motion are
obtained for the system considered in the given space from the
least action principle. Yet we will follow another way and try
to build physical space and major properties of elementary
objects in this space from dynamics of a finite set of special
discrete objects (so called {\it byuons}) \cite{4,5}. Note that the
development of physical
comprehension of elementary processes on the base of modern
superstring models \cite{6}${}^-$\cite{11},
 unfortunately, also gives no evidences
for how structured is the observed space itself which is
obtained, according to one of the models, by means of
compactification of six dimensions in a ten-dimensional space.
New theoretical approaches found in construction of physical
space have given the chance to look in a new way also at the
most studied object of the classic and quantum field theory, the
electromagnetic field.  In the present paper we will not touch
upon many problems widely covered in the literature and
concerned with the photon and electromagnetic field: whether the
photon has a rest mass (Proc's equation), whether or not it
possesses a longitudinal component of field, being artificially
removed by the Lorentz's condition \cite{12}, etc. We try only to
explain the origin of the Maxwell's equations as a peculiar
change in the structure of physical space due to dynamics of
byuons.  Chronologically, the author has arrived to the theory
of byuons when investigating a system of spinor and boson fields
interacting with the electromagnetic field \cite{13}${}^-$\cite{17}.
 As a generalized coordinate in this system, the velocity of
propagation of interactions $c = c(x,t)$ was taken where $x$ is the
coordinate of a certain one-dimensional compact space $M_1$, $t$ is
time, and the electric charge $e$ of the fields is assumed to be
some implicit function of $c(x,t)$, i.e. $e = e [c(x,t)]$.  Thus the
author anticipated the local violation of gauge invariance and
Lorentz's group extended on translation (Poincaret's group) in
order to gain an insight into the process of physical space
formation.
\vskip20pt

\section{\bf  Fundamental Theoretical Concepts of Physical Space
Origin.} \vskip10pt

In Ref. \cite{4,5}, the conception of formation of the observed
physical space $R_3$ from a finite set of byuons is given. The byuons
${\bf\Phi}(i)$ are one-dimensional vectors and have the form:
$${\bf\Phi}(i) ={\bf A_g}x(i),$$
where $x(i)$ is the "length" of the byuon, a real (positive, or negative)
value depending on the index $i=0,1,2,...,k...$, a quantum number of
${\bf\Phi}(i)$; under $x(i)$ a certain time charge of the byuon may be
meant
(with $x(i)$ in centimeters).  The vector ${\bf A}_g$  represents the
cosmological vectorial potential, a new basic vectorial constant
\cite{4,5,16,17}. It takes only two values:
$$ {\bf A}_g = \left\{ A_g\atop-\sqrt{-1} A_g\right\} ,$$
where $A_g$ is the modulus of the cosmological vectorial potential
($ A_g \approx 1.95\times 10^{11}$ CGSE units).  According to the theory
\cite{4,5}, the
value $A_g$ is the limiting one. In reality, there exists in
nature, in the vicinity of the Earth, a certain summary
potential, i.e. the vectorial potential fields from the Sun
($A_\odot \approx 10^8$ CGSE units), the Galaxy ($ \sim 10^{11}$ CGSE
units), and the
Metagalaxy ($ > 10^{11}$ CGSE units) are superimposed on the constant $A_g$
resulting probably in some turning of ${\bf A}_\Sigma$  relative to the
vector
${\bf A}_g$ in the space $R_3$ and in a decrease of it.

Hence in the theory
of physical space (vacuum) which the present article leans upon,
the field of the vectorial potential introduced even by Maxwell
gains a fundamental character. As is known, this field was
believed as an abstraction. All the existing theories are
usually gauge invariant. For example, in classical and
quantum electrodynamics, the vectorial potential is defined with
an accuracy of an arbitrary function gradient, and the scalar
one is with that of time derivative of the same function, and
one takes only the fields of derivatives of these potentials, i.
e. magnetic flux density and electric field strength, as real.

ln Refs.\cite{13}${}^-$\cite{17},
local violation of the gauge invariance and
Poincaret's group was supposed, and the elementary particle
charge and quantum number formation processes were investigated
in some set, therefore the potentials gained an unambiguous
physical meaning there. In the present paper, this is a finite
set of byuons.

The works by D.Bohm and Ya.Aharonov \cite{18}
discussing the special meaning of potentials in quantum
mechanics are the most close to the approach under
consideration, they are confirmed by numerous experiments
\cite{19}.

For understanding the byuon dynamics, present an
extract from the author's materials of investigations on metrics
in the one-dimensional compact space $M_I$ \cite{14}.

Let us assume  that the velocity $c$ in the space $M_I(x,t)$ is a function
of $x$ and
$t$, i.e. $c = c(x,t)$.  Then the metrics of such a manifold may be
represented as
$$d[c(x,t)t]^2 - dx^2 = 0.\eqno{(1)}$$
Rewrite Eq. (1) in the form
$$\{d[c(x,t) t] - dx\}\{ d[c(x,t) t] + dx\} = 0. \eqno{(2)}$$
Removing brackets in Eq.(2) and dividing it by $dt$  yield two equations
$$ t\frac{\partial c}{\partial t} + c + t\frac{\partial c}{\partial x}
\frac{dx}{dt}  - \frac{dx}{dt}  = 0, \eqno{(3.1)}$$
$$ t\frac{\partial c}{\partial t} + c + t\frac{\partial c}{\partial x}
\frac{dx}{dt}  + \frac{dx}{dt}  = 0. \eqno{(3.2)}$$

At limiting points of the compact space $M_I$  the differentials
$dx$ and $dt$ are not independent, therefore we may assume $\frac{dx}{dt} =
c$.
Then we have from Eq.(3)
$$ (3.1) \Rightarrow  \frac{\partial c}{\partial t} + \frac{\partial
c}{\partial x}c = 0, \eqno{(4)} $$
$$ (3.2) \Rightarrow  t\frac{\partial c}{\partial t} + t\frac{\partial
c}{\partial x}c + 2c = 0. \eqno{(5)}$$
Eq.(4) with boundary conditions $t = t_o , x = x_o$ has a
solution $c(x,t) = const = c_o$, which is in accord with a postulate
of special theory of relativity.

Seek a solution of Eq.(5) with the same boundary conditions as for Eq.(4).
The system of characteristic equations may be written as \cite{14}:
$$ \frac{dc}{dt} = \frac{\partial c}{\partial t} + c\frac{\partial
c}{\partial x},   \eqno{(6)}$$
$$\frac{dx}{dt} = c, \eqno{(7)}$$
$$ \frac{d(\frac{\partial c}{\partial x})}{dt} = - \frac{\partial
c}{\partial x}\left(\frac{\partial c}{\partial x} +
\frac{2}t\right),\eqno{(8)}$$
$$ \frac{d(\frac{\partial c}{\partial t})}{dt} = - \left[\frac{\partial
c}{\partial t}\left(\frac{\partial c}{\partial x} + \frac{2}t\right) -
\frac{2c}{t^2}\right].\eqno{(9)}$$
Solving Eq. (8), we find
$$c(x,t) = \frac{x}{t(t/t^* - 1)} + f(t).\eqno{(10)}$$
Here $t^*$ is constant of integration, $f(t)$ an arbitrary function of $t$.
Setting $f(t) = 0$ in Eq.(10) gives
$$c(x,t) = \frac{x}{t(t/t^* - 1)}.\eqno{(11)}$$
The function $c(x,t)$ satisfies (5) and at a fixed $x_0$ has the form shown
in Fig. 1.
As is seen, $c(x,t)$ has discontinuities at $t = t^*$ and $t = 0$.
Since a signal from point $x_1$ to point $x_2$  in $M_I$  cannot be
transmitted instantaneously (with the
infinite speed), we assume that only the limiting values $c = \pm c_0$ can
be realized in nature.  One may come to this conclusion
when considering non-classical variation calculus, too. It is
known from Ref. \cite{24} that if restrictions are imposed on a
variable function minimizing a functional (in our case, $c(x,t)$
may be thought of as imposed on by inequalities $-c_0 \le c(x,t) \le c_o$
due to physical vacuum properties), then, from the viewpoint of
minimum functional, only boundary values of the function being
varied should be optimum.


\begin{figure}[thb]  
\centerline{\psfig{figure=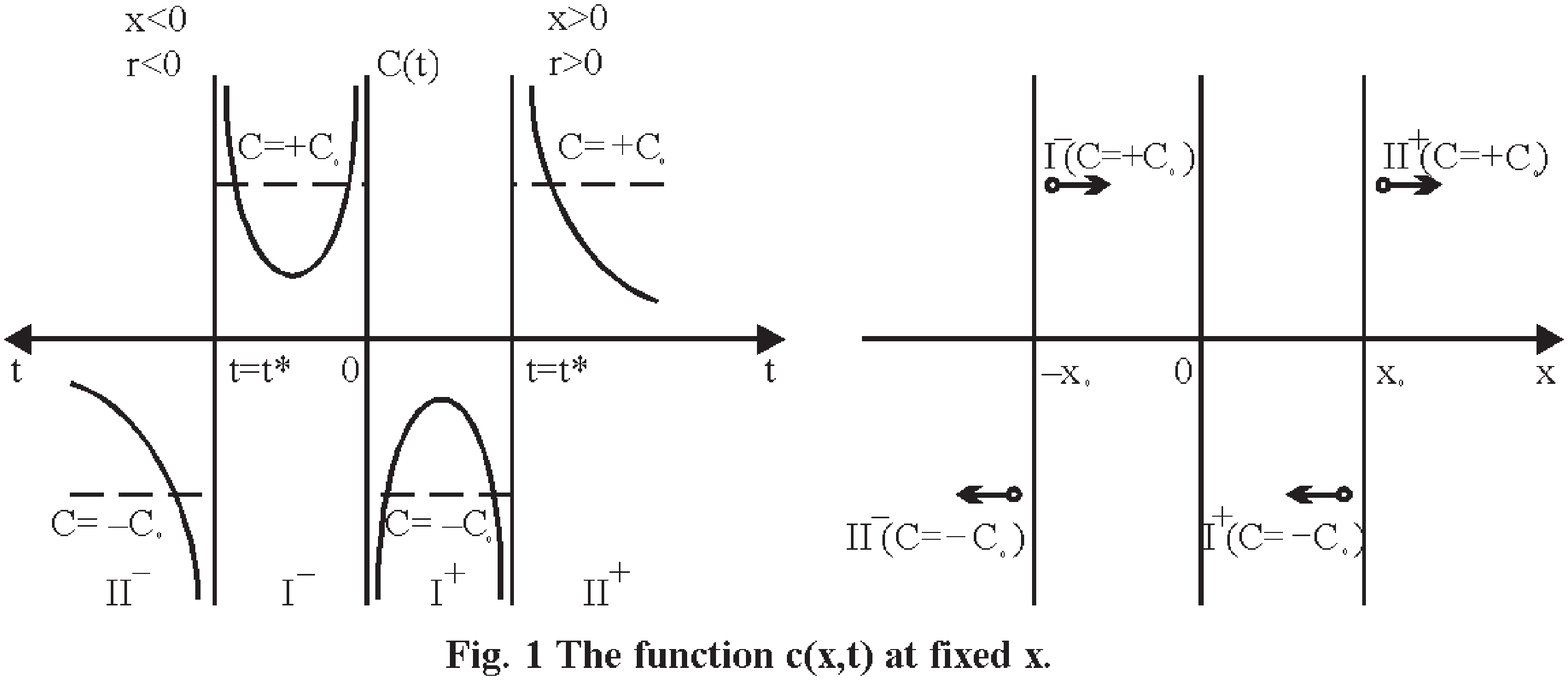,angle=0,height=60mm}}
\vspace{5mm}
\caption{The function $c(x,t)$ at fixed x.}
\end{figure}

The byuons may be in four  vacuum states (VS)
$II^+, I^+, II^-, I^-,$ in which they discretely change
the value ot $x(i)$: the state $II^+$ discretely increases ($c = c_0 =
\tilde x_0/\tau_0$,
where $\tilde x_0$ - quantum of space ($\approx 10^{-33}$cm), $\tau_0$ -
quantum
of time ($\approx 10^{-43}c$)) and $I^+$ decreases $x(i)>0$ ($c = -c_o =
-\tilde x_0/\tau_0$);
the states $II^-$ and $I^-$ discretely increase
or decrease the modulus of $x(i)<0$, respectively ($II^-$ corresponds to $c
= - c_0$,
$I^-$ corresponds to $c = c_0$). The sequence of discrete changes of $x(i)$
value is defined as a proper discrete time of the byuon.
The byuon vacuum states originate randomly \cite{4,5}.

In Refs. \cite{4,5}, the following hypothesis has been put forward:
\vskip10pt
{\it It is suggested that the space $R_3$ observed is built up as a result
of
minimizing the potential energy of byuon interactions in the
one-dimensional space $R_1$ formed by them. More precisely, the
space $R_3$ is fixed by us as a result of dynamics arisen of
byuons. The dynamic processes and, as a consequence, wave
properties of elementary particles appear therewith in the space
$R_3$ for objects with the residual positive potential energy of byuon
interaction (objects observed). }
\vskip10pt
Let us briefly list the results obtained earlier  when investigating the
present model of
physical vacuum (see Appendix):

1. The existence of a new long-range
interaction in nature, arising when acting on physical vacuum by
the vectorial potential of high-current systems, has been
predicted \cite{25,28}.

2. All the existing interactions (strong,
weak, electromagnetic and gravitational ones) along with the new
interaction predicted have been qualitatively explained in the
unified context of changing in three periods of byuon
interactions with characteristic scales - $x_0 = \tilde x_0 k \approx
10^{-17}$cm,
$ct^* = \tilde x_0 kN \approx 10^{-13}$cm, and $\tilde x_0 kNP \approx
10^{28}$cm,
determined from the minimum potential energy of byuon interaction
\cite{4,5}.

3. Masses of leptons, basic barions and mesons have been found
\cite{16,17}.

4. The constants of weak interaction (vectorial and axial ones)
and of strong interaction have been calculated \cite{16,17}.

5. The origin of the galactic and intergalactic magnetic fields has
been explained as a result of existence of an insignificant
($\approx 10^{-15}$) asymmetry in the formation process of $R_3$ from the
one-dimensional space of byuons \cite{4,5}.

6. The matter density observed in the Universe ($\approx 10^{-29} g/cm^3$)
has been computated \cite{4,5}.

7. The origin of the relic radiation has been explained on the
basis of unified mechanism of the space $R_3$ formation from
one-dimensional space $R_1$ of byuons \cite{4,5}, etc.

Let us explain item 1 briefly. It is shown in Ref. \cite{17} that masses of
all
elementary particles are proportional to the modulus of ${\bf A}_g$ (see
Appendix). If
now we direct the vectorial potential of a magnetic system in
some space region towards the vector ${\bf A}_g$ then any material body
will be forced out of the region where $|{\bf A}_\Sigma | < |{\bf A}_g|$ .
The new force is nonlocal, nonlinear, and represented by a complex
series in $\Delta A$, a difference in changes of $|{\bf A}_\Sigma|$ due to
the
potential of a current at location points of a sensor and a test
body \cite{5,28}. This force is directed mainly by the direction of the
vector ${\bf A}_g$,
but as the latest experiments have shown, there is also an
isotropic component of the new force in natures, which component
acts  omnidirectionally from the space of the maximum decrease
in $|{\bf A}_\Sigma|$. Corresponding to it are probably the even terms in a
series representing the new force \cite{5,28}.

One of the important predictions of the theory is revealing a new
information channel
in the Universe which is associated with the existence of a
minimum object with positive potential energy, so called object
$4b$, arising in the minimum four-contact interaction of byuons in
the vacuum states $II^+, I^+, II^-, I^-$. In four-contact byuon
interaction, a minimum action equal to $h$ (Planck's constant)
occurs, and the spin of the object appears. Hence the greater
part of the potential energy of byuon interaction is transformed
into spin of the object $4b$. The residual (after minimization)
potential energy of the object $4b$ is equal to $\approx 33eV$. It is
identified with the rest mass of this object in the space $R_1$.
In agreement with Refs. \cite{4,5}, the indicated minimum object $4b$
has, according to Heisenberg uncertainty relation, the
uncertainty in coordinate $\Delta x\approx 10^{28}$cm in $R_3$.
The total energy of these objects determines near $98\%$ energy of the
Universe as
well as its matter density observed.  In Refs. \cite{4,5} it is shown
that the existence of the $4b$-objects interpreted by the author
as pairs of electron-type neutrino and electron-type
antineutrino ($\nu_e \Leftrightarrow \tilde \nu_e$), causes a finiteness
of the velocity of propagation of interaction c being an infinite
quantity at $t = 0$  and  $t = t^*$ according to Eq. (11).
\vskip20pt

\section{\bf On The Nature Of The Electromagnetic Field.}
\vskip10pt

Within the limits of an article it is difficult to unravel the
problem as a whole, i.e. to develop a physical and mathematical
model of the phenomenon. Therefore we dwell basically on a new
physical model of the electromagnetic field resulting from the
above presented model of physical space. But, as is known, the
new is very often contained in the forgotten old if that is
looked at in a new way.

Consider the Weyl's geometrical model of the electromagnetic field
\cite{29}.
Historically, the Weyl's geometry was proposed for explaining by curvature
of space not
only the long-range gravitational field as in the theory of
Einstein but another long-range field, the electromagnetic
field, too.

Gravity is well explained by the Einstein's theory
in terms of space curvature. This suggests that the
electromagnetic field also can be attributed to a certain space
property instead of considering this field as simply "immersed"
in space. Thus, some more general space than the Riemannean
space underlying the Einstein's theory, had to be proposed to
describe the existence of either gravitational or
electromagnetic forces and to unify the long-range interactions.

The curvature of space required by the Einstein's theory can be
represented in terms of parallel displacement of a vector moving
along a closed contour, which leads to the effect that the final
direction of the vector will not be the same as the former. The
Weyl's generalization was in assuming that the final vector has
not only a different direction but a different length, too (and
this is the most important thing). In the Weyl's geometry, there
is no absolute procedure of comparing elements of length at two
different points if these are not infinitely close to each
other.

The comparison may be made only relative to a line
segment connecting the two points, and different ways will give
different results as to ratio of the two elements of length. In
order to have mathematical theory of lengths one must
arbitrarily establish a length standard at each point and then
relate any length at anyone point to the local standard for this
point. Then we will have a fixed value of the vector length at
any point but this value varies with the local standard of
length.

Let's consider a vector of length $l$ positioned at a point
with coordinates $x^\mu (\mu = 1, 2, 3, 4)$.  Assume that this vector
is transferred to the point $x^\mu + \delta x^\mu$. The change in its
length
$\delta l$ will be proportional to $l$ and $\delta x^\mu$ so that
$\delta l = l k_\mu \delta x^\mu$ where $k_\mu$ is additional parameters of
the field appearing in the
theory together with the Einsteinean $g_{\mu\nu}$ and being equally
fundamental.

Assume that the length standards have been changed
so that lengths are multiplied by a coefficient $\lambda (x)$. Then $l$
goes into $l' = l\lambda (x)$ and $l + \delta l$ does into
$$l'+ \delta l' = (l + \delta l)\lambda(x + \delta x) = (l + \delta
l)\lambda (x)
+ l \frac{\partial\lambda}{\partial x^\mu}\delta x^\mu.$$
Hence
$$\delta l' = \lambda\delta l + l\frac{\partial\lambda}{\partial
x^\mu}\delta x^\mu
= \lambda l(k_\mu + \frac{\partial\lg\lambda}{\partial x^\mu})\delta
x^\mu.$$
Thus, $\delta l' = l' k'_\mu\delta x^\mu$  where
$$k'_\mu = k_\mu + \frac{\partial\lg\lambda}{\partial x^\mu}\eqno{(12)}$$
If our vector is transferred by parallel displacement
 along a closed contour, a change in its length will be $\delta l =
l\Omega_{\mu\nu}
\delta S^{\mu\nu}$  where
$$\Omega_{\mu\nu} = \frac{\partial k_\mu}{\partial x^\nu} - \frac{\partial
k_\nu}{\partial x^\nu},$$
and $\delta S^{\mu\nu}$ notes an element of area limited by a small
contour.

Let us show that the new field characteristic $k_\mu$  appearing in the Weyl's
theory have the meaning of electromagnetic potentials. As is
known \cite{12,31}, the antisymmetric tensor $F_{\mu\nu}$ of the
electromagnetic field has the form $F_{\mu\nu} = \partial_\mu A_\nu -
\partial_\nu A_\mu$
where $A_\mu$ and $A_\nu$ are components of the real covariant
four-vectorial electromagnetic potential $A_\alpha (\alpha = 0,1,2,3)$.
Comparing the $\Omega_{\mu\nu}$ and $F_{\mu\nu}$ tensors we see that,
indeed, the
additional parameters characterizing proportionality of change
in length of the vector $l$ while going from the point $x^\mu$ to the
point $x^\mu + \delta x^\mu$, have the meaning of electromagnetic
potentials.
They go through the transformations (12) corresponding not to a
change in geometry but only to a change in choosing artificial
length standards. The derivatives $\Omega_{\mu\nu}$ have a geometrical
meaning without regard to standards of length and correspond to
physically significant parameters, electric and magnetic fields.
Thus, the geometry of Weyl assures just what is necessary for
describing the gravitational and electromagnetic fields in terms
of geometry.

Hence when assuming a mathematical algorithm of
transition of events from $R_1$ to $R_3$  as having been found, the
electromagnetic and gravitational interaction of objects can be
assured by changing period of byuon interactions $k, N$  and scale
lengths $x_0$ and $ct^*$, respectively. The electromagnetic field is
caused therewith by changing $|{\bf A}_\Sigma |$ due to some scalar $A_0$,
and
magnetic field is due to a vectorial potential ${\bf A}$ but with the
complete return of the vector ${\bf A}$ to the initial point of the $R_3$ -
space while tracing some closed contour l (i.e. in the absence
of spacial anisotropy). Note once more that the electromagnetic
constant of interaction (fine structure constant) $e^2/hc_0$ will
be constant in all reference  systems (see Appendix).

Let's now answer the question "At the cost of which VSs of byuons is the
electromagnetic field (i.e. variation of $l$ at the cost of $x_0$ and
$ct^*$ variations) described ?" There can be two such variants of
byuons VSs for describing the electromagnetic field (waves,
photons). These are the pairs of byuons in VS $II^+I^-$  and pairs
in VS $II^-I^+$.  The later is explained by the fact that, for
example, VSs $II^+$ and $I^-$ have, firstly, the same $c = c_0$, and
secondly, they propagate in the same direction of the $R_1$-space
(similarly for VS $II^-I^+$).  Explain the former. If byuon VSs have
the same $c$, then, in accordance with the ideology of charge
formation \cite{15}, the quantum transitions (correlation) between
them cannot lead to formation of electric charge since to do
this, the presence of VSs with opposite magnitudes of c is quite
necessary for locking information within some local region of
the space $R_1$ and hence $R_3$.
\vskip20pt

\section{\bf Conclusion.}
\vskip10pt

Thus, it is qualitatively shown in the article
that at the cost of varying periods of byuon interaction, it is
possible to describe the electromagnetic field basing on the
Weyl's geometrical theory of this field.
\vskip20pt

\appendix{\bf }
\vskip10pt

In \cite{16,17} by using an unordinary varying the action for the system of
spinor and
boson fields interacting with electromagnetic one, taking $c(x,t)$ as a
generalized
coordinate, considering $e=e[c(x,t)]$, the following formal results are
obtained:

1. Expressions for $h$ and leptons are found\footnote{It is necessary to
explain that in 1984-86 the author with Yu.N.Babaev
had not yet known that into the expressions for the lepton masses $(A.2)$
not the
mass of neutrino, but that of pair "neutrino-antineutrino" enters (so
called 4b-object).}:

$$h = \frac{(|{\bf A}| x_0)^2}{c}\times\frac{x_0}{ct^*},\eqno{(A.1)}$$
$$m_e = m_{\nu_e} \frac{ct^*}{x_0},\hspace{1cm}m_\mu = m_{\nu_\mu}
\frac{ct^*}{2x_0},\hspace{1cm}m_\tau = m_{\nu_\tau}
\frac{ct^*}{8x_0},\eqno{(A.2)}$$
$$m_{\nu_\mu}c^2 = |{\bf A}| A_0
\frac{2x_0^3}{(ct^*)^2},\hspace{1cm}m_{\nu_\tau}c^2 = |{\bf A}| A_{0_\tau}
\frac{24\times8x_0^3}{(ct^*)^2},\eqno{(A.3)}$$
$$A_0e = \frac{1}2(|{\bf A}| x_0)\times\left(\frac{x_0}{ct^*}\right)^2\times|{\bf
A}|,\hspace{1cm}A_0 =
\frac{2\sqrt{3}e}{\sqrt{x_0ct^*}},\hspace{1cm}A_{0_\tau} =
\frac{2\sqrt{3}e}{\sqrt{2x_0ct^*}}.\eqno{(A.4)}$$
From these relationships three important conclusions follows:
$$|{\bf A}|x_0 = \frac{(hc)^{3/2}}{4\sqrt{3}e^2} = const_1,\eqno{(A.5)}$$
$$\frac{x_0}{ct^*} = \left(\frac{4\sqrt{3}e^2}{hc}\right)^2 =
const_2,\eqno{(A.6)}$$
$$e^2 = \frac{(|{\bf A}|
x_0)^2}{4\sqrt{3}}\times\left(\frac{x_0}{ct^*}\right)^{3/2}.\eqno{(A.7)}$$

2. The expressions for the Planck's constant $(A.1)$ and value of electric
charge $(A.7)$
are invariant relative to variation of the parameters $|{\bf A}|$, $x_0$,
$ct^*$
in the relationships $(A.5)$ and $(A.6)$. Hence, when varying, for example,
$x_0$ and $ct^*$ so that the ratio $(A.6)$ is always equal
to $const_2$, the expressions for $h$ $(A.1)$ and $e$ $(A.7)$ will remain
unchanged. Similar is the situation with the parameters $|{\bf A}|$ and
$x_0$ in $(A.5)$.
Thus, we arrive to a conclusion that in nature there are possible a certain
set of
earlier unknown objects determined by the product ${\bf A}\times x_0$ which
our world is based
on, since its fundamental properties, particularly electromagnetic ones
(the
fine structure constant ${e^2}/{hc}$) determined by such constants as $h,
e, c$, remain
unchanged on this set.

3. Expressions for axial and vectorial constants of weak interactions,
are deduced\footnote{Here and below all numerical estimations are given
at $c = c_0, e = e_0, {\bf A} = {\bf A_g}$.}~:
$$c_A = \frac{1}2e_0^2x_0ct^* \approx 1,4333\times 10^{-49} erg\cdot
cm^3,$$
$$c_V = 2x_0^3\left(e_0|{\bf A}|+\frac{hc}{4x_0}\right) \approx
1,4336\times 10^{-49} erg\cdot cm^3.$$

4. Constant of strong interactions for proton: $g^2_p = \frac{1}{6}$, for
$\pi^0$-meson: $g_{\pi^0}^2 = \frac{4}{6}$ and expressions for masses of
these
particles, are obtained:
\begin{eqnarray}
m_p c^2 &\approx& \frac{\sqrt{hc}}{2}
\left(\frac{4\sqrt{3}e^2}{hc}\right)^3\times
\left\{1 - \frac{1}{2}\sqrt{\frac{3x_0}{ct^*}}\right\}\times\nonumber\\
&& \qquad\qquad\left[\sqrt{1 + \frac{1}{6}\left(1 -
\frac{1}{2}\sqrt{\frac{3x_0}{ct^*}}\right)^{-2}}
+ 1\right]\times|{\bf A}|\approx 924 MeV,\nonumber
\end{eqnarray}
\begin{eqnarray}
m_{\pi_0}c^2 &\approx& \frac{\sqrt{hc}}{2}
\left(\frac{4\sqrt{3}e^2}{hc}\right)^3\times
\left\{1 - 2\sqrt{\frac{3x_0}{ct^*}}\right\}\times\nonumber\\
&&\qquad\qquad\left[\sqrt{1 + \frac{1}{6}\left(1 -
2\sqrt{\frac{3x_0}{ct^*}}\right)^{-2}} - 1\right]\times|{\bf A}|\approx
132 MeV.\nonumber
\end{eqnarray}
Also as it is shown in Ref. \cite{17}, masses of all
leptons are proportional to the modulus of ${\bf A}$, too:
\begin{eqnarray}
m_e c^2 &=&
\frac{\sqrt{hc}}{2k}\left(\frac{hc}{4\sqrt{3}e^2}\right)^3\times|{\bf
A}|,\,\,
m_{\nu_e}c^2 =
\frac{\sqrt{hc}}{2k}\left(\frac{hc}{4\sqrt{3}e^2}\right)\times|{\bf
A}|\fill2cm;\nonumber\\
m_\mu c^2 &=&
3^{1/4}\sqrt{hc}\left(\frac{4\sqrt{3}e^2}{hc}\right)^{3,5}\times|{\bf
A}|,\,\,
m_{\nu_\mu} c^2 =
2\times3^{1/4}\sqrt{hc}\left(\frac{4\sqrt{3}e^2}{hc}\right)^{5,5}\times|{\bf
A}|;\nonumber\\
m_\tau c^2 &=&
12\sqrt{2}\times3^{1/4}\sqrt{hc}\left(\frac{4\sqrt{3}e^2}{hc}\right)^{3,5}\times|{\bf
A}|,\,\,\nonumber\\
&&\qquad\qquad\qquad m_{\nu_\tau} c^2 =
96\sqrt{2}\times3^{1/4}\sqrt{hc}\left(\frac{4\sqrt{3}e^2}{hc}\right)^{5,5}\times|{\bf
A}|.\nonumber
\end{eqnarray}

5. The size of space quantum $\tilde x_0$ (without introducing the
gravitation force),
is found:
$$\tilde x_0 = 2\sqrt{3}\left(\frac{4\sqrt{3}e^2}{hc}\right)^9\times
\frac{hc}{m_ec^2}\left(\frac{m_e}{m_\mu}\right)^2 \approx
2,77986\times10^{-33}cm.$$

6. The electric, barion, and lepton charges, as well as strangeness, are
defined.
Note that masses of all leptons, $p$ and $\pi^0$-mesons are expressed in
terms of the
single constant, modulus of a certain vectorial potential ${\bf A}$.
Note once more that the size of space quantum is found without introducing
gravitational force. Both conclusions are very important as showing that
the nature of
gravitation may be explained by way of developing physical views.
With the use of a known fact of proportionality of space quantum to
$\sqrt{\frac{Gh}{c^3}}$, where $G$ is the gravitational constant,
one may show that
$$ \frac{\sqrt{G}}{c^2} =
\left(\frac{hc}{4\sqrt{3}e^2}\right)\times(0,6861k|{\bf A}|)^{-1}.$$

7. The value of magnetic flux density $B$ in $R_3$-space can be represented
as $\frac{1}{16}\times\frac{|{\bf A_g}|\tilde x_0}{4\pi
x_0^2}\times\frac{\tilde x_0}{x_0}$
where $\frac{\tilde x_0}{x_0}$ is sine of the angle between the byuon and
the plane
tangent to the sphere of radius $x_0$ at the point of byuon coming out.
Estimations show that $B \approx 5\times 10^{-6} Gs$ \cite{4,5}.

\nonumsection{References}

\end{document}